\documentclass[aps,pra,twocolumn,amsfonts,amssymb,amsmath,showpacs,
floatfix,nofootinbib,groupedaddress,superscriptaddress,citesort]{revtex4}%
\usepackage{mathrsfs}
\usepackage{amsfonts}
\usepackage{amstext}
\usepackage{amsmath}
\usepackage{amssymb}
\usepackage{color}
\usepackage{bm}
\usepackage{bbm}
\usepackage[dvips]{graphicx}
\def\qed{\leavevmode\unskip\penalty9999 \hbox{}\nobreak\hfill
     \quad\hbox{\leavevmode  \hbox to.77778em{%
              \hfil\vrule   \vbox to.675em%
               {\hrule width.6em\vfil\hrule}\vrule\hfil}}
     \par\vskip3pt}

\begin{document}

\preprint{APS/123-QED}
\title{Quantum correlation exists in any non-product state\\}

\author{Yu Guo}
\affiliation{School of Mathematics and Computer Science, Shanxi Datong University, Datong, Shanxi 037009, China}%

\author{Shengjun Wu}
\affiliation{Kuang Yaming Honors School, Nanjing University, Nanjing, Jiangsu 210093, China}

\begin{abstract}

Simultaneous existence of correlation in complementary bases is a
fundamental feature of quantum correlation, and we show that this
characteristic is present in any non-product bipartite state. We propose
a measure via mutually unbiased bases to study this feature of
quantum correlation, and compare it with other measures of
quantum correlation for several families of bipartite states.

\end{abstract}

\pacs{03.65.Ud, 03.65.Db, 03.65.Yz.}
\maketitle


Quantum systems can be correlated in ways
inaccessible to classical objects.
This quantum feature of correlations
not only is the key to our understanding
of quantum world, but also is essential for the
powerful applications of quantum information and quantum
computation \cite{EPR,Sc,EinsteintoBornletter,
BBPS96,Wer89,horodecki09rev,
Usha08,Skrzypczyk14,Vedral14,Georgescu13,Strobel14,
Yu2009,Dakic12,wu2009correl,Luo2008,
Modi2011}.
In order to characterize the correlation in quantum state,
many approaches have been proposed to reveal different
aspects of quantum correlation, such as the
various measures of entanglement \cite{horodecki09rev,Usha08}
and the various measures of discord and
related measures \cite{wu2009correl,Luo2008,Modi2011,
OZ01,Luo2011}, etc.
It is believed that some aspects of quantum correlation
could still exist without the presence of
entanglement and these aspects could be revealed via local
measurements with respect to some basis of a local system.

The simultaneous existence
of complementary correlations in
different bases is revealed very early
by the Bell's inequalities \cite{Bellinequality}.
Bell's inequalities quantify quantum correlation via
expectation values of local complementary observables.
In \cite{Wu2014}, the feature of
genuine quantum correlation is revealed by
defining measures based on invariance under a basis change:
for a bipartite quantum state, the classical correlation is
the maximal correlation present in a certain optimum basis,
while the quantum correlation is characterized as a series
of residual correlations in the bases
mutually unbiased (MU) to the optimum basis.
In this paper, we use the fact that the essential feature of the quantum
correlation is that it can be present in any
two mutually unbiased bases (MUBs) simultaneously.
Thus, one of the two bases is not necessarily the
optimum basis to reveal the maximal classical correlation in this paper.
With respect to the measure proposed here, we shall show that
only the product states do not contain quantum correlation.
A product state contains neither any quantum correlation nor
any classical correlation; while any non-product bipartite state contains
correlation that is fundamentally quantum!
We shall also reveal interesting properties of this measure by
comparing this measure to other measures of quantum
correlation for several families of bipartite states.

The MUBs constitute now a basic ingredient in many applications
of quantum information processing: quantum state
tomography \cite{Wootters1989},
quantum cryptography \cite{Bechmann2000},
discrete Wigner function \cite{Wootters1987},
quantum teleportation \cite{Koniorczyk2001},
quantum error correction codes \cite{Paz2005},
and the mean king's problem \cite{Englert2001}.
Two orthonormal bases $\{|\psi_i\rangle\}$ and
$\{|\phi_j\rangle\}$ of a $d$-dimensional Hilbert space $H$ are said
to be mutually unbiased
if and only if
\begin{equation}
|\langle\psi_i|\phi_j\rangle|
=\frac{1}{\sqrt{d}},\quad \forall\ 1\leq i,j\leq d.
\end{equation}
In a $d$-dimensional Hilbert space, there exist at least $3$ MUBs
(when $d$ is a power of a prime number, a full set of $d +1$ MUBs exists,
more details can be found in \cite{Ivanovic81}).

We recall the quantity defined in \cite{Wu2014}.
Let $H_{ab}=H_a\otimes H_b$ with $\dim H_a=d_a$
and $\dim H_b=d_b$ be the state space of
the bipartite system A+B shared by Alice and Bob.
Let $\{|i\rangle\}$ and $|j'\rangle$ be the
orthonormal bases of $H_a$ and $H_b$ respectively.
Alice selects a basis
$\{ |i\rangle\}$ of
$H_a$ and performs
a measurement projecting her system onto the basis states.
The Holevo quantity $\chi\{\rho_{ab}|\{|i\rangle\}\}$
of $\rho_{ab}$ with respect to Alice's local
projective measurement onto the basis $\{|i\rangle\langle i|\}$,
is defined as
$\chi\{\rho_{ab}|\{\left|i\right\rangle\}\}
=\chi\{p_{i};\rho_{i}^{b}\}\equiv S(\sum_{i}p_{i}\rho_{i}^{b})
-\sum_{i}p_{i}S(\rho_{i}^{b})$. A basis
$\{\left|i\right\rangle\}$ that achieves the maximum
(denoted as $C_{1}(\rho_{ab})$)  of the
Holevo quantity is called a $\chi$-basis of $\rho_{ab}$.
There could exist
many $\chi$-bases for a state $\rho_{ab}$, and the set
of these bases is denoted as  $\Gamma_{\rho_{ab}}$.
Let $\Omega_{\Pi^a}$ be the set of all bases that are mutually
unbiased to $\Pi^a$, $\Pi^a\in\Gamma_{\rho_{ab}}$.
The quantity of
quantum correlation in \cite{Wu2014}, denoted by
$Q_{2}(\rho_{ab})$, is defined as
\begin{equation}
Q_{2}(\rho_{ab})\equiv\max_{\Pi^a\in\Gamma_{\rho_{ab}}}\max_{\tilde{\Pi}^a\in\Omega_{\Pi^a}}
\chi\{\rho_{ab}|\tilde{\Pi}^a\}. \label{defQ2}
\end{equation}
In other words,
$Q_2$ is defined as the Holevo quantity of Bob's accessible
information about Alice's results, maximized over
Alice's projective measurements in the bases that are mutually
unbiased to a $\chi$-basis $\Gamma_{\rho_{ab}}$,
and further maximized over all possible $\chi$-bases (if not unique).

{\it Correlation measure based on MUBs.---}We now present
our approach in a more general way.
Let $\Delta$ denote the set of all two-MUB sets, i.e.,
\begin{eqnarray*}
\Delta=\{\{\{|i_1\rangle\},\{|j_2\rangle\}\}:\{|i_1\rangle\}\
\mbox{\rm is MU to}\ \{|j_2\rangle\}\}.
\end{eqnarray*}
We define
\begin{equation}
\mathcal{C}(\rho_{ab}) \equiv
\max_{(\Pi^a_{1},\Pi^a_{2})\in\Delta}
\min\{ \chi\{\rho_{ab}|\Pi^a_{1}\},\chi\{\rho_{ab}|\Pi^a_{2}\}\} . \label{defC}
\end{equation}
The quantity $\mathcal{C}$ represents the maximal amount of
correlation that is present simultaneously in two MUBs.
In a sense, $\mathcal{C}$ is more essential than $Q_2$
since the maximum in the former one
is taken over arbitrarily two MUBs.
Thus, $\mathcal{C}$ may reveal more quantum correlation than $Q_2$.
Similar to the other usual measures of quantum correlation,
$\mathcal{C}$ is local unitary invariant,
that is,
$\mathcal{C}(\rho_{ab})
=\mathcal{C}(U_a\otimes U_b\rho_{ab}U_a^\dag\otimes U_b^\dag)$ for
any unitary operators $U_a$ and $U_b$
acting on $H_a$ and $H_b$ respectively.

{\it The nullity of $\mathcal{C}$.---}
Now we show that any bipartite quantum state contains nonzero correlation
simultaneously in two mutually unbiased bases unless it
is a product state, this result is stated as the following theorem.

\if Now we show that any bipartite quantum state contains nonzero correlation
simultaneously in two mutually unbiased bases unless it
is a product state, this result is stated as the following theorem, while the proof
is left to the Appendix.
\fi

{\it Theorem.} $\mathcal{C}(\rho_{ab})=0$ if
and only if $\rho_{ab}$ is a product state.

{\it Proof.}\quad The `if' part is obvious, and we only need to show
the `only if' part. In other words, we only need to
prove that $\rho_{ab}=\rho_a\otimes\rho_b$ if either $\chi(\rho_{ab}|\Pi_1^a)=0$ or
$\chi(\rho_{ab}|\Pi_2^a)=0$ for any MUB pair $(\Pi_1^a,\Pi_2^a)\in\Delta$.
It is equivalent to show that both $\chi(\rho_{ab}|\Pi_1^a) \neq 0$ and
$\chi(\rho_{ab}|\Pi_2^a) \neq 0$ for a certain MUB pair
$(\Pi_1^a,\Pi_2^a)\in\Delta$ if $\rho_{ab}$ is not a product state.

We assume that $\rho_{ab}$ is not a product state, then the maximal
classical correlation is nonzero, i.e., $C_1(\rho_{ab})\neq0$.
Let $\{|e_i\rangle\}\in\Gamma_{\rho_{ab}}$, we have
$\chi(\rho_{ab}|\{|e_i\rangle\} ) \neq 0$. Therefore, we only need to find a
second basis (MU to $\{|e_i\rangle\}$) such that the
corresponding Holevo quantity is nonzero.
We denote the projective measurement corresponding to $\{|e_i\rangle\}$ by
$\prod=\{\prod_k=|e_k\rangle\langle e_k|\}$.
Then
$\prod(\rho_{ab}) =\sum_k\prod_k\otimes I_b\rho_{ab}\prod_k\otimes I_b
=\sum_kp_k^{(1)}|e_k\rangle\langle e_k|\otimes\rho_k^{b(1)}$.
As $C_1(\rho_{ab})\neq0$, we know that $\rho_{k_0}^{b(1)}\neq \rho_b$
and $\rho_{l_0}^{b(1)}\neq\rho_b$ at least for some $k_0$ and $l_0$.
We arbitrarily choose a basis $\{|f_i\rangle\}$ that is MU to $\{|e_i\rangle\}$.
If $\chi\{\rho_{ab}|\{|f_i\rangle\}\} \neq 0$, then we already obtain
the second basis and the theorem is true.

If $\chi\{\rho_{ab}|\{|f_i\rangle\}\}=0$, we can construct
the MUB pair as follows. As in this case, the measurement
corresponding to $\{|f_i\rangle\}$ yields the following output state
\begin{eqnarray*}
\left(\begin{array}{cccc}
p_1^{(2)}\rho_b&0&\cdots&0 \\
0&p_2^{(2)}\rho_b&\cdots&0 \\
\vdots&\vdots&\ddots&\vdots\\
0&0&\cdots&p_d^{(2)}\rho_b\end{array}\right).
\end{eqnarray*}
Thus, $\rho_{ab}$ can be represented as
\begin{eqnarray*}
\left(\begin{array}{cccc}
p_1^{(2)}\rho_b&*&\cdots&* \\
*&p_2^{(2)}\rho_b&\cdots&* \\
\vdots&\vdots&\ddots&\vdots\\
*&*&\cdots&p_d^{(2)}\rho_b\end{array}\right)
\end{eqnarray*}
with respect to the local basis $\{|f_i\rangle\}$, and
at least one of the off-diagonal blocks is not zero
(otherwise, $\rho_{ab}$ is a product state).
Without loss of generality we assume that
the (1,2)-block-entry of the above matrix is nonzero.
It follows that there exists a 2 by 2 unitary matrix $U_2$, such that,
under the local basis $\{U_2\oplus I_{d-2}|f_i\rangle\}$,
the state admits the form
\begin{eqnarray*}
\left(\begin{array}{cccc}
q_1^{(2)}\varrho_b&*&\cdots&* \\
*&q_2^{(2)}\sigma_b&\cdots&* \\
\vdots&\vdots&\ddots&\vdots\\
*&*&\cdots&p_d^{(2)}\rho_b\end{array}\right)
\end{eqnarray*}
with $\varrho_b\neq\rho_b$ and $\sigma_b \neq\rho_b$.
That is $\chi\{\rho_{ab}|\{U_2\oplus I_{d-2}|f_i\rangle\}\}\neq0$.
This unitary matrix $U_2$ can be chosen as
\begin{eqnarray*}
U_2=\left(\begin{array}{cc}
\sqrt{1-\epsilon^2}&\epsilon \\
-\epsilon&\sqrt{1-\epsilon^2}\end{array}\right).
\end{eqnarray*}
with $\epsilon$ a very small positive number. Even though
$\chi\{\rho_{ab}|\{U_2\oplus I_{d-2}|f_i\rangle\}\}$
could be very small, it is nonzero.
As $\epsilon$ is a very small and
$\chi\{\rho_{ab}|\{|e_i\rangle\}\}\neq0$, we also have
$\chi\{\rho_{ab}|\{U_2\oplus I_{d-2}|e_i\rangle\}\}\neq0$.
Thus, the Holevo quantity is nonzero at least for a certain MUB pair
(i.e., $\{U_2\oplus I_{d-2}|e_i\rangle\}$ and
$\{U_2\oplus I_{d-2}|f_i\rangle\}$),
and therefore $\mathcal{C}(\rho_{ab})\neq0$.

Thus, $\mathcal{C}(\rho_{ab})\neq0$ for
any $\rho_{ab}$ that is not a product state.
The proof is completed.\hfill$\blacksquare$

In a sense, this theorem implies that, any non-product bipartite state contains
genuine quantum correlation, and $\mathcal{C}$ reveals the amount
of quantum correlation in the state.
In addition, we know that $\mathcal{C}$ is
different from the quantity $Q_2$ in \cite{Wu2014}
since $Q_2(\rho_{cq})=0$ for any classical-quantum state
$\rho_{cq}$ while $\mathcal{C}=0$ only for product states.
The difference between the measure $\mathcal{C}$ and other
measures of quantum correlation shall be discussed below for several
families of bipartite states in more details.

{\it Examples.---}
Now, we shall calculate the quantity for several families of
bipartite states, and see how our measure in terms of MUBs is
well justified as a measure of quantum correlation.

For a bipartite
pure state with the Schmidt decomposition
$|\psi\rangle=\sum_{i}\sqrt{\lambda_{i}}|a_{i}\rangle |b_{i}\rangle$,
$\mathcal{C}=Q_2=S(\rho_{B})=S(\rho_{B})
=\sum_{i}-\lambda_{i}\log_{2}\lambda_{i}$.
It can be easily checked that $\mathcal{C}$ coincides with
the entropy of either reduced state for any pure state, which is also
the usual measure of entanglement in a pure state.

Next, we consider the Werner states of a $d\otimes d$ dimensional
system \cite{Wer89},
\begin{equation}
\rho_{w}=\frac{1}{d(d-\alpha)}(I-\alpha P) ,
\label{Wernerstates}
\end{equation}
where $-1\leq\alpha\leq1$, $I$ is the
identity operator in the $d^2$-dimensional
Hilbert space, and
$P=\sum_{i,j=1}^{d}|i\rangle \langle j|\otimes|j\rangle \langle i|$
is the operator that exchanges A and B.
For a local measurement with respect to
basis states $\left\{ |e_i\rangle \right\} $ of $H_a$,
with probability $p_{i}=\frac{1}{d}$, Alice will obtain the $k$-th
basis state $|e_k\rangle $, and Bob will be left with the
state $\rho_{k}^{b}=\frac{1}{d-\alpha}(I-\alpha|e_k'\rangle\langle e_k'|)$,
where $|e_k'\rangle=\sum_j\alpha_{kj}|j'\rangle$
with $\alpha_{kj}=\langle e_k|j\rangle$.
It is straightforward to show that
\begin{equation}
\mathcal{C}(\rho_w)=\chi\left\{ p_{i};\rho_{i}^{B}\right\}
=\log_{2}(\frac{d}{d-\alpha})+\frac{1-\alpha}{d-\alpha}\log_{2}(1-\alpha).
\end{equation}
The entanglement of formation $E_{f}$ for the Werner states is given
as
$E_{f}(\rho_{w})=h\left(\frac{1}{2}(1+\sqrt{1-[\max(0,\frac{d\alpha-1}{d-\alpha})]^{2}})\right)$,
with $h(x)\equiv - x\log_{2} x- (1-x) \log_{2}(1-x)$ \cite{wootters01}. The
three different measures of quantum correlation, i.e.,
$\mathcal{C}$, the quantum discord $D$ and
the entanglement of formation $E_f$, are illustrated in Fig. 1
for comparison. From this figure, we see that the curve for entanglement of
formation intersects the other two curves; thus, $E_f$
can be larger or smaller than $\mathcal{C}$.

\begin{figure}[h]
\centering \includegraphics[width=8.5cm]{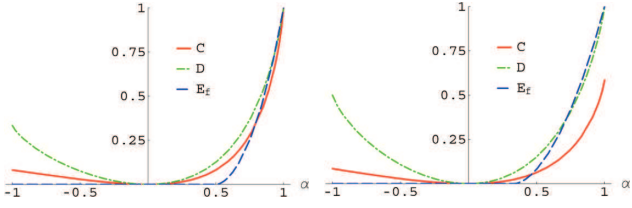}
\caption{(color online). Measures of quantum correlation for the Werner states as functions
of $\alpha$ when $d=2$ (left) and $d=3$ (right). The red curve
represents our measure $\mathcal{C}$, the green curve represents
the quantum discord $D$ and the blue curve represents the entanglement
of formation $E_{f}$.}
\label{fig1}
\end{figure}

For the $d\otimes d$ isotropic states
\begin{equation}
\rho=\frac{1}{d^2-1}((1-\beta)I+(d^2\beta-1)P^+),\quad \beta\in[0,1],
\label{Isotropicstates1}
\end{equation}
where $P^+=|\Phi^+\rangle\langle\Phi^+|$,
$|\Phi^+\rangle=\frac{1}{\sqrt{d}}\sum_i|i\rangle|i'\rangle$
is the maximally entangled pure state in $\mathbb{C}^d\otimes \mathbb{C}^d$.
Let $\{|e_k\rangle\langle e_k|\}$ be an arbitrarily
given projective measurement on Alice's part.
Bob's state after after Alice gets the k-th measurement result is
\begin{equation*}
\rho_{k}^b=\frac{1}{d^2-1}(d(1-\beta)I+(d^2\beta-1)|e_k'\rangle\langle e_k'|),
\label{Isotropicstates2}
\end{equation*}
where $|e_k'\rangle=\sum_j\alpha_{kj}|j'\rangle$
with $\alpha_{kj}=\langle e_k|j\rangle$.
As the eigenvalues of $\rho_{k}^b$ does not depend on the
basis for Alice's measurement, one can easily show that
\begin{eqnarray}
\mathcal{C}(\rho)&=&\log_2d+\frac{d\beta+1}{d+1}\log_2\frac{d\beta+1}{d+1}\nonumber\\
&&+\frac{d-d\beta}{d+1}\log_2\frac{d-d\beta}{d^2-1}.
\label{Isotropicstates3}
\end{eqnarray}
The
entanglement of formation $E_{f}$ for the isotropic states is given
as  \cite{terhal,Fei2006}
\begin{equation*}
E_{f}(\rho)
=\left\{\begin{array}{ll}
0, & \beta\leq\frac{1}{d},\\
h(\gamma)+(1-\gamma)\log_2(d-1), &\frac{1}{d}<\beta<\frac{4(d-1)}{d^2},\\
\frac{(\beta-1)d\log_2(d-1)}{d-2}+\log_2d, &\frac{4(d-1)}{d^2}\leq \beta\leq 1,
\end{array}\right.
\label{Isotropicstates4}
\end{equation*}
where $\gamma=\frac{1}{d}(\sqrt{\beta}+\sqrt{(d-1)(1-\beta)})^2$.
The quantum discord of the isotropic state is ~\cite{chitambar}
\begin{eqnarray*}
D(\rho)&=&\beta\log_2\beta+\frac{1-\beta}{d+1}\log_2\frac{1-\beta}{d^2-1}\nonumber\\
&&-\frac{1+d\beta}{d+1}\log_2\frac{1-\beta-\frac{1}{d}+d\beta}{d^2-1}.
\label{Isotropicstates5}
\end{eqnarray*}
The three different measures of quantum correlation,
i.e., $\mathcal{C}$, the quantum discord $D$ and
the entanglement of formation $E_f$, are illustrated in Fig. 2
for comparison. From this figure,
we see that the curve for entanglement of
formation intersects the other two curves;
thus, $E_f$ can be larger or smaller than $\mathcal{C}$.

\begin{figure}[h]
\centering \includegraphics[width=8.5cm]{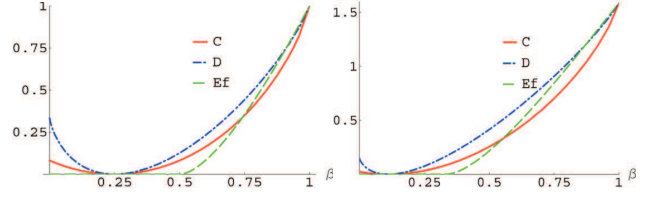}
\caption{(color online). Measures of quantum correlation for the isotropic states as functions
of $\beta$ when $d=2$ (left) and $d=3$ (right). The red curve
represents our measure $\mathcal{C}$, the green curve represents
the quantum discord $D$ and the blue curve represents the entanglement
of formation $E_{f}$.}
\label{fig2}
\end{figure}

As the last example, we consider a family of two-qubit states that are
equivalent to Bell-diagonal states under local unitary transformations.
This family of states admit the form
\begin{equation}
\sigma_{ab}=\frac{1}{4} (I_2\otimes I_2
+\sum_{j=1}^{3}r_{j}\sigma_{j}\otimes\sigma_{j} ).
\label{twoqubitsymmtricstatessimplified}
\end{equation}
We rearrange the three numbers $\{r_1,r_2,r_3\}$ according to their
absolute values and denote the rearranged set as
$\{ \overline{r}_1, \overline{r}_2, \overline{r}_3\}$ such that
$|\overline{r}_1| \geq |\overline{r}_2| \geq |\overline{r}_3|$.
Next we show that
\begin{equation}
\mathcal{C}(\sigma_{ab})=1-h(\frac{1+\sqrt{(r_1^2+r_2^2)/2}}{2}).
\end{equation}
A projective measurement performed on qubit A can be written as
$P_{\pm}^{a}=\frac{1}{2}(I_2 \pm\vec{n}\cdot\vec{\sigma})$,
parameterized by the unit vector $\vec{n}$.
When Alice obtains $p_{\pm}$, Bob will be in the corresponding states
$\rho_{\pm}^{b}=\frac{1}{2}(I_2 \pm\sum_{j}n_{j}r_{j}\sigma_{j})$,
each occurring with probability $\frac{1}{2}$. The entropy $S(\rho_{\pm}^{b})$
reaches its minimum value $h(\frac{1+|r_{1}|}{2})$
when $\vec{n}=(1,0,0)$.
Let $\vec{n_1}=(x,y,0)$ and $\vec{n_2}=(a,b,0)$ with $ax+by=0$,
then
$P^{(1)}_{\pm}$ is mutually unbiased to $P^{(2)}_{\pm}$,
where
$P^{(1)}_{\pm}=\frac{1}{2}(I_2\pm \vec{n_1}\cdot\vec{\sigma})$,
$P^{(2)}_{\pm}=\frac{1}{2}(I_2\pm \vec{n_2}\cdot\vec{\sigma})$.
It is immediate that
$\chi\{\sigma_{ab}|P^{(1)}_{\pm}\}=1-h(\frac{1+\sqrt{x^2r_1^2+y^2r_2^2}}{2})$
and $\chi\{\sigma_{ab}|P^{(2)}_{\pm}\}=1-h(\frac{1+\sqrt{a^2r_1^2+b^2r_2^2}}{2})$.
Thus $\mathcal{C}(\sigma_{ab})=1-h(\frac{1+\sqrt{(r_1^2+r_2^2)/2}}{2})$ as desired since
$h(c)$ is a monotonic decreasing function when $c\geq\frac{1}{2}$.
Our quantity $\mathcal{C}$ is compared with the quantum discord $D$ and
the entanglement of formation $E_f$ for $\rho_1$ and $\rho_2$ in Fig. 3.

\begin{figure}[h]
\centering \includegraphics[width=8.5cm]{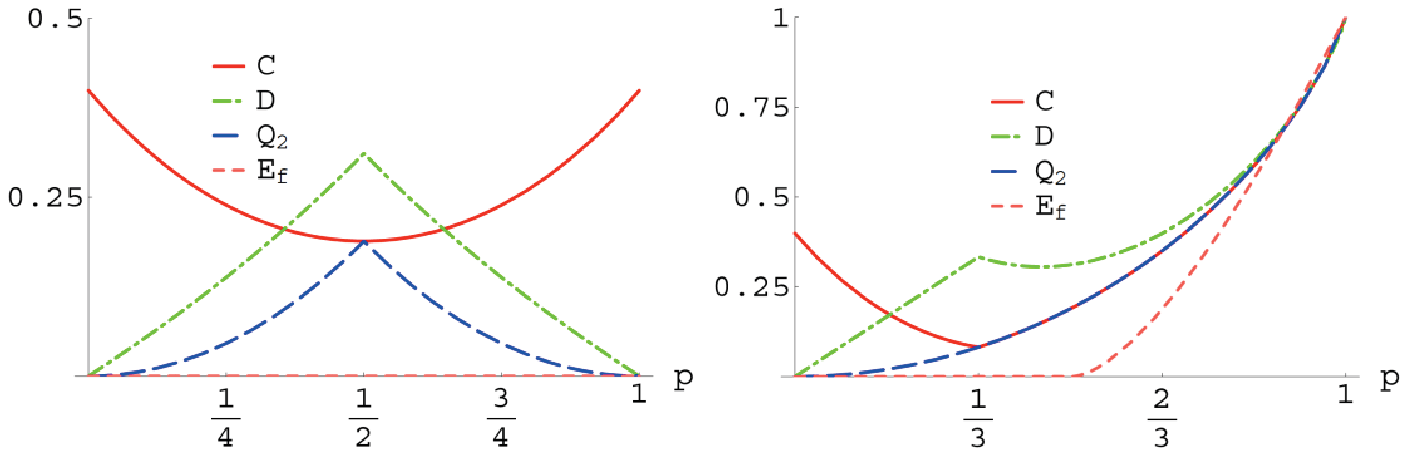}
\caption{(color online). Different measures of quantum correlation for two special classes of states:
$\rho_1 = \frac{1}{2} \left| \psi^+ \right\rangle \left\langle \psi^+ \right|
+\frac{p}{2} \left| \phi^+ \right\rangle \left\langle \phi^+ \right|
+\frac{1-p}{2} \left| \phi^- \right\rangle \left\langle \phi^- \right| $ (left) and
$\rho_2 = p \left| \psi^- \right\rangle \left\langle \psi^- \right|
+\frac{1-p}{2} \left( \left| \psi^+ \right\rangle \left\langle \psi^+ \right|
+ \left| \phi^+ \right\rangle \left\langle \phi^+ \right| \right)$ (right).
In each figure, the red curve
represents our measure $\mathcal{C}$, the green curve represents
the quantum discord $D$, the blue curve represents
the measure $Q_2$, and the blown curve represents the entanglement
of formation $E_{f}$.}
\label{fig3}
\end{figure}

From the left figure of Fig. 3, it is clear that $\mathcal{C}$ is quite different from both
$D$ and $Q_2$.  We have $\mathcal{C}(\rho_1)<D(\rho_1)$ when $p$ is closed to $\frac{1}{2}$,
while $\mathcal{C}(\rho_1)>D(\rho_1)$ when $p$ is closed to $0$ or $1$; we also have
$\mathcal{C}(\rho_1)=Q_2(\rho_1)$ when $p=\frac{1}{2}$,
and $\mathcal{C}(\rho_1)$ increases monotonously while $Q_2(\rho_1)$
decreases  monotonously when $p$ deviates from $\frac{1}{2}$.
In Fig. 3, the difference between our measure $\mathcal{C}$ and the
other measures is well illustrated by the extreme cases when $p=0$ or $1$ in the
left figure and when $p=0$  in the right figure. For example, for
$\sigma = \frac{1}{2} \left| \psi^+ \right\rangle \left\langle \psi^+
\right| +\frac{1}{2} \left| \phi^+ \right\rangle \left\langle \phi^+ \right|$,
our measure has a finite value while the other measures vanish.

{\it Correlation revealed via more MUBs.---}
In addition, we can define a quantity based on $m$ MUBs ($3\leq m\leq \dim H_a+1$), namely,
\begin{eqnarray}
\mathcal{C}_m(\rho_{ab})& \equiv&
\max_{(\Pi^a_{1},\Pi^a_{2},\cdots,\Pi^a_{m})\in\Delta_m}
\min\{ \chi\{\rho_{ab}|\Pi^a_{1}\},\nonumber\\
&&~~~\chi\{\rho_{ab}|\Pi^a_{2}\},\cdots,\chi\{\rho_{ab}|\Pi^a_{m}\} .
\end{eqnarray}
where
\begin{eqnarray*}
\Delta_m&=&\{(\Pi^a_{1},\Pi^a_{2},\cdots,\Pi^a_{m}):\\
&&~~~\Pi^a_{k}\ \mbox{\rm is MU to}\ \Pi^a_l\ \mbox{for any }\ k \neq l\}.
\end{eqnarray*}
It is clear that $\mathcal{C}_{k+1}\leq \mathcal{C}_{k}\leq \mathcal{C}$.
The following are obvious from the arguments in the previous examples:
i) $\mathcal{C}_m(\rho)=0$ if and only if $\rho$ is a product state, ii)
$\mathcal{C}_m=\mathcal{C}$ for both Werner states
and the isotropic states, and iii)
$\mathcal{C}_3(\sigma_{ab})=1-h(\frac{1+\sqrt{(r_2^2+r_3^2)/2}}{2})$
for the family of two-qubit
states in Eq.~(\ref{twoqubitsymmtricstatessimplified}).

In conclusion, we have provided a very different approach to quantify
quantum correlation in a bipartite quantum state.
Our approach captures the essential feature of
quantum correlation: the simultaneous existence of correlations
in complementary bases. We have proved that the only states that don't have this feature are
the product states, which contains no correlation (classical or quantum)
at all. Thus, any non-product state contains correlation that is fundamentally quantum.
This feature of quantum correlation characterized here could be the key feature that enables
quantum key distribution (QKD) with entangled states, since the
quantum correlation that exists simultaneously in  MUBs, which can be
quantified by $\mathcal{C}$, is the resource for entanglement-based QKD via
MUBs.

\smallskip

Y. Guo is supported by the Natural Science Foundation of
China (No. 11301312, No. 11171249),
the Natural Science Foundation of Shanxi
(No. 2013021001-1,  No. 2012011001-2)
and the Research start-up fund for Doctors of Shanxi Datong University
(No. 2011-B-01).
S. Wu is supported by the Natural Science Foundation of China (No. 11275181).
This paper is dedicated to professor Jinchuan Hou for his sixtieth birthday.



\end{document}